\documentclass[letter,onecolumn,romanappendices]{ieeeconf}
\let\proof\relax   

\usepackage{mathpple}
\usepackage{times}
\usepackage{amsthm,xpatch}
\usepackage{amsmath,amsfonts}
\usepackage{xfrac}
\usepackage{cite}
\usepackage{amssymb}
\usepackage{dsfont}
\usepackage{graphicx, subfigure}
\usepackage{color}
\usepackage{breqn}
\usepackage{mathtools}
\usepackage{bbm}
\usepackage{latexsym}
\usepackage[ruled, linesnumbered]{algorithm2e}
\usepackage{accents}
\usepackage{rotating}
\usepackage{tikz}
\usepackage[mathscr]{euscript}
\usepackage{bm}
\usepackage[frak=lucida]{mathalpha} 
\usepackage{comment}
\usepackage{balance}
\usepackage{optidef}
\usepackage{float}

\newtheorem{theorem}{Theorem}
\newtheorem{remark}{Remark}

\newtheorem*{example*}{Example}
\DeclareMathOperator*{\argmax}{\arg\!\max}

  \renewcommand{\baselinestretch}{0.98}

\makeatletter
\newcommand*{\transpose}{%
  {\mathpalette\@transpose{}}%
}
\newcommand*{\@transpose}[2]{%
  \raisebox{\depth}{$\m@th#1\intercal$}%
}
\makeatother

\IEEEoverridecommandlockouts

\begin{document}

\makeatletter
\newcommand{\raisemath}[1]{\mathpalette{\raisem@th{#1}}}
\newcommand{\raisem@th}[3]{\raisebox{#1}{$#2#3$}}
\makeatother

\newcommand{\mstk}{\hspace{-0.145cm}*}

\newcommand{\mstl}{\hspace{-0.105cm}*}

\newcommand{\mstm}{\hspace{-0.175cm}*}

\newcommand{\SB}[3]{
\sum_{#2 \in #1}\biggl|\overline{X}_{#2}\biggr| #3
\biggl|\bigcap_{#2 \notin #1}\overline{X}_{#2}\biggr|
}

\newcommand{\Mod}[1]{\ (\textup{mod}\ #1)}

\newcommand{\overbar}[1]{\mkern 0mu\overline{\mkern-0mu#1\mkern-8.5mu}\mkern 6mu}

\makeatletter
\newcommand*\nss[3]{%
  \begingroup
  \setbox0\hbox{$\m@th\scriptstyle\cramped{#2}$}%
  \setbox2\hbox{$\m@th\scriptstyle#3$}%
  \dimen@=\fontdimen8\textfont3
  \multiply\dimen@ by 4             
  \advance \dimen@ by \ht0
  \advance \dimen@ by -\fontdimen17\textfont2
  \@tempdima=\fontdimen5\textfont2  
  \multiply\@tempdima by 4
  \divide  \@tempdima by 5          
  \ifdim\dimen@<\@tempdima
    \ht0=0pt                        
    \@tempdima=\fontdimen5\textfont2
    \divide\@tempdima by 4          
    \advance \dimen@ by -\@tempdima 
    \ifdim\dimen@>0pt
      \@tempdima=\dp2
      \advance\@tempdima by \dimen@
      \dp2=\@tempdima
    \fi
  \fi
  #1_{\box0}^{\box2}%
  \endgroup
  }
\makeatother

\makeatletter
\renewenvironment{proof}[1][\proofname]{\par
  \pushQED{\qed}%
  \normalfont \topsep6\p@\@plus6\p@\relax
  \trivlist
  \item[\hskip\labelsep
        \itshape
    #1\@addpunct{:}]\ignorespaces
}{%
  \popQED\endtrivlist\@endpefalse
}
\makeatother

\makeatletter
\newsavebox\myboxA
\newsavebox\myboxB
\newlength\mylenA

\newcommand*\xoverline[2][0.75]{%
    \sbox{\myboxA}{$\m@th#2$}%
    \setbox\myboxB\null
    \ht\myboxB=\ht\myboxA%
    \dp\myboxB=\dp\myboxA%
    \wd\myboxB=#1\wd\myboxA
    \sbox\myboxB{$\m@th\overline{\copy\myboxB}$}
    \setlength\mylenA{\the\wd\myboxA}
    \addtolength\mylenA{-\the\wd\myboxB}%
    \ifdim\wd\myboxB<\wd\myboxA%
       \rlap{\hskip 0.5\mylenA\usebox\myboxB}{\usebox\myboxA}%
    \else
        \hskip -0.5\mylenA\rlap{\usebox\myboxA}{\hskip 0.5\mylenA\usebox\myboxB}%
    \fi}
\makeatother

\xpatchcmd{\proof}{\hskip\labelsep}{\hskip3.75\labelsep}{}{}

\pagestyle{plain}

\title{\fontsize{21}{28}\selectfont Multi-Message Private Information Retrieval:\\ A Scalar Linear Solution}

\author{Ningze Wang, Anoosheh Heidarzadeh, and Alex Sprintson\thanks{The authors are with the Department of Electrical and Computer Engineering, Texas A\&M University, College Station, TX 77843 USA (E-mail: \{ningzewang, anoosheh, spalex\}@tamu.edu).}
}

\maketitle 

\thispagestyle{plain}

\begin{abstract}
In recent years, the Multi-message Private Information Retrieval (MPIR) problem has received significant attention from the research community. In this problem, a user wants to privately retrieve $D$ messages out of $K$ messages whose identical copies are stored on $N$ remote servers, while maximizing the download rate. The MPIR schemes can find applications in many practical scenarios and can serve as an important building block for private computation and private machine learning applications. The existing solutions for MPIR  require a large degree of subpacketization, which can result in large overheads, high complexity, and impose constraints on the system parameters. These factors can limit practical applications of the existing solutions. In this paper, we present a methodology for the design of scalar-linear MPIR schemes. Such schemes are easy to implement in practical systems as they do not require partitioning of messages into smaller size sub-messages and do not impose any constraints on the minimum required size of the messages. Focusing on the case of $N=D+1$, we show that when $D$ divides $K$, our scheme achieves the capacity, where the capacity is defined as the maximum achievable download rate. When the divisibility condition does not hold, the performance of our scheme is the same or within a small additive margin compared to the best known scheme that requires a high degree of subpacketization. 
\end{abstract}

\section{Introduction}
In the Private Information Retrieval (PIR) problem \cite{CGKS1995}, a user wishes to retrieve one or more messages belonging to a dataset---by downloading the minimum possible amount of information from one or more remote servers that store copies of the dataset, while revealing no information about the identities of the desired messages to the servers. 

Several variations of the PIR problem have been studied by the research community.
This includes 
multi-server single-message PIR 
\cite{SJ2017,TER2017,TGKHHER2017,SJ2018,BU18,TSC2019,BAWU2020,VBU2020,LTFH2021,SS2021,BAU2021,ZTSP2021ISIT,LJJ2021}, 
multi-server multi-message PIR~\cite{BU2018}, 
single-server single-message PIR with side information 
\cite{KGHERS2017No0,KGHERS2020,HKS2019Journal,HKS2018,HKS2019,KHSO2019,KHSO2021,HS2021,HS2022Reuse,LJ2022,GLH2022arXiv}, 
multi-server single-message PIR with side information 
\cite{T2017,WBU2018,WBU2018No2,KKHS12019,KKHS22019,KGHERS2020,CWJ2020,LG2020CISS,KH2021}, 
and single-server and multi-server multi-message PIR with side information 
\cite{HKGRS2018,LG2018,SSM2018,HKRS2019,KKHS32019,HS2022LinCap}.  

In this work, we revisit the problem of multi-server multi-message PIR (MPIR)~\cite{BU2018} focusing on \emph{scalar-linear} solutions.  
In MPIR, identical copies of a dataset of ${K>1}$ messages are stored on ${N>1}$ non-colluding servers, and a user is interested in retrieving $D>1$ (out of $K$) messages belonging to the dataset, 
while hiding the identities of these $D$ messages from each of the servers. 
The need to retrieve multiple messages can be motivated by several practical applications. 
For example, consider a setting in which we need to train a machine learning algorithm on a subset of the data stored at a remote server, such that the identify of the required data is kept private. 
As another example, in a financial application, there might be a need to compute a function of certain financial indicators, such as prices of certain stocks or bonds, while keeping the identities of the stocks or bonds private. 

The capacity of MPIR, defined as the maximum achievable download rate over all MPIR schemes, was characterized previously in~\cite{BU2018}, for a wide range of parameters $N,K,D$. 
While the existing solutions for MPIR are of a great theoretical interest, 
their practical applicability is limited due to high degree of subpacketization. 
Specifically, in order to achieve the capacity, the solution presented in~\cite{BU2018} requires to divide each message into $N^2$ or more  sub-messages (or sub-packets), where $N$ is the number of servers. 
The high level of subpacketization  presents several performance and implementation challenges. 
First, the high subpacketization limits the range of system parameters in which the PIR can operate. 
For example, in the case that the subpacketization level depends on the number of servers, the limit on the size of messages imposes a limit on the number of servers. 
Also, a high level of subpacketization can increase the complexity of the private information retrieval process due to the need to maintain and manage a large amount of meta-data. 

Motivated by the scheme due to Tian \emph{et al.}~\cite{TSC2019} for the single-message PIR setting, in this work we present a scalar-linear scheme for the multi-message PIR (MPIR) setting. 
Our scheme does not require a division of the messages into smaller-size sub-messages and does not impose any constraint on the minimum size of the messages. 
Our scheme is designed for the case when $N=D+1$ and achieves the capacity when $D\mid K$. 
This implies that our scheme has the same performance as the capacity-achieving scheme due to Banawan and Ulukus~\cite{BU2018} that requires a significant degree of subpacketization. 
When ${D\nmid K}$ and ${D<\frac{K}{2}}$, the rate achievable by our scheme is no less than that of~\cite{BU2018}, and 
surprisingly, in some cases, our scheme can achieve a higher rate. 
When ${D\nmid K}$ and ${D>\frac{K}{2}}$, our scheme achieves a rate which is slightly lower than that of~\cite{BU2018}. 



\section{Problem Setup}\label{sec:SN}
We denote random variables and their realizations by bold-face and regular symbols, respectively. 
For any integer $i\geq 1$, we denote $\{1,\dots,i\}$ by $[i]$. 
Let $q$ be a prime power, and let $m\geq 1$ be an integer. 
Let $\mathbbmss{F}_q$ be a finite field of order $q$, ${\mathbbmss{F}_q^{\times} := \mathbbmss{F}_q\setminus \{0\}}$ be the multiplicative group of $\mathbbmss{F}_q$, and $\mathbbmss{F}_{q}^{m}$ be the vector space of dimension $m$ over $\mathbbmss{F}_q$.

Let $N$ and $K$ be two arbitrary integers such that $N,K>1$. 
Consider $N$ non-colluding servers each of which stores an identical copy of $K$ messages ${\mathrm{X}_1,\dots,\mathrm{X}_K}$, where $\mathrm{X}_i\in \mathbbmss{F}_q^{m}$ for $i\in [K]$ is a vector of length $m$ with entries in $\mathbbmss{F}_q$. 
For every ${\mathrm{S}\subset [K]}$, we denote $\{\mathrm{X}_i: i\in \mathrm{S}\}$ by $\mathrm{X}_{\mathrm{S}}$. 

Let $D$ be an arbitrary integer such that ${1<D\leq K}$, and let $\mathbbmss{W}$ be the set of all $D$-subsets of $[K]$.  
Consider a user who wishes to retrieve the set of $D$ messages $\mathrm{X}_{\mathrm{W}}$ for a given $\mathrm{W}\in \mathbbmss{W}$. 
We refer to $\mathrm{X}_{\mathrm{W}}$ as \emph{demand}, $\mathrm{W}$ as the \emph{demand's index set}, 
and $D$ as the \emph{demand's size}. 

We assume that: (i) $\mathbf{X}_1,\dots,\mathbf{X}_K$ are independent and uniformly distributed over $\mathbbmss{F}_{q}^{m}$. 
That is, ${H(\mathbf{X}_i) = B:=m\log_2 q}$ for all ${i\in [K]}$, and ${H(\mathbf{X}_{\mathrm{S}})= |\mathrm{S}| B}$ for all ${\mathrm{S}\subseteq [K]}$; 
(ii) $\mathbf{X}_{1},\dots,\mathbf{X}_K$ and $\mathbf{W}$ are independent; 
(iii) $\mathbf{W}$ is uniformly distributed over $\mathbbmss{W}$; and
(iv) the distribution of $\mathbf{W}$ is initially known by each server, whereas the realization $\mathrm{W}$ is not initially known by any server.

Given $\mathrm{W}$, the user generates a query $\mathrm{Q}_n^{[\mathrm{W}]}$ for each ${n\in [N]}$, and sends it to server $n$. 
Each query $\mathrm{Q}_n^{[\mathrm{W}]}$ is a deterministic or stochastic function of $\mathrm{W}$, and independent of $\mathrm{X}_1,\dots,\mathrm{X}_K$. 
For each $n\in [N]$, the query $\mathrm{Q}_n^{[\mathrm{W}]}$ must not reveal any information about the demand's index set $\mathrm{W}$ to server $n$. 
That is,
for every ${{\mathrm{W}^{*}}\in \mathbbm{W}}$, it must hold that
\begin{equation*}
\mathbb{P}(\mathbf{W}={\mathrm{W}^{*}}|\mathbf{Q}_n^{[\mathbf{W}]}=\mathrm{Q}_n^{[\mathrm{W}]})=\mathbb{P}(\mathbf{W}={\mathrm{W}^{*}}) \quad \forall n\in [N]. 
\end{equation*}
We refer to this condition as the \emph{privacy condition}. 

Upon receiving $\mathrm{Q}_n^{[\mathrm{W}]}$, server $n$ generates an answer $\mathrm{A}_n^{[\mathrm{W}]}$, and sends it back to the user. 
The answer $\mathrm{A}_n^{[\mathrm{W}]}$ is a deterministic function of $\mathrm{Q}_n^{[\mathrm{W}]}$ and $\mathrm{X}_1,\dots,\mathrm{X}_K$.
That is, ${H(\mathbf{A}_n^{[\mathrm{W}]}|\mathbf{Q}_n^{[\mathrm{W}]},\mathbf{X}_1,\dots,\mathbf{X}_K)=0}$. 
The user must be able to recover their demand $\mathrm{X}_{\mathrm{W}}$ given the collection of answers
${\mathrm{A}^{[\mathrm{W}]}:=\{\mathrm{A}_n^{[\mathrm{W}]}: n\in [N]\}}$, the collection of queries ${\mathrm{Q}^{[\mathrm{W}]}:=\{\mathrm{Q}_n^{[\mathrm{W}]}: n\in [N]\}}$, and the realization $\mathrm{W}$. 
That is,
$H(\mathbf{X}_{\mathrm{W}}| \mathbf{A}^{[\mathrm{W}]},\mathbf{Q}^{[\mathrm{W}]})=0$. 
We refer to this condition as the \emph{recoverability condition}.

The problem is to design a protocol for generating a collection of queries $\mathrm{Q}^{[\mathrm{W}]}$ and the corresponding collection of answers $\mathrm{A}^{[\mathrm{W}]}$ for any given $\mathrm{W}$ 
such that both the privacy and recoverability conditions are satisfied.
This problem, which was originally introduced in~\cite{BU2018}, is referred to as \emph{Multi-message Private Information Retrieval (MPIR)}. 

In contrast to~\cite{BU2018} which considers the space of all MPIR protocols, 
in this work we focus on \emph{scalar-linear MPIR protocols}, which we refer to as \emph{linear MPIR protocols} for short, in which each server's answer to the user's query consists only of (scalar-) linear combinations of the messages. 
That is, for each $n\in [N]$, the answer $\mathrm{A}^{[\mathrm{W}]}_n$ consists of one or more linear combinations of the messages $\mathrm{X}_1,\dots,\mathrm{X}_K$ with combination coefficients from $\mathbbmss{F}_q$.  
In addition, for each $n\in [N]$, the query $\mathrm{Q}^{[\mathrm{W}]}_n$ can be represented by the support and the nonzero combination coefficients pertaining to the linear combinations that constitute the answer $\mathrm{A}^{[\mathrm{W}]}_n$. 

We define the \emph{rate} of a protocol as the ratio of the number of bits required by the user, 
i.e., ${H(\mathbf{X}_{\mathrm{W}})=DB}$, to the expected number of bits downloaded from all servers, 
i.e., ${\sum_{n=1}^{N} \mathbb{E}[L(\mathbf{A}^{[\mathrm{W}]}_n)]}$, 
where $L(\mathbf{A}^{[\mathrm{W}]}_n)$ is the number of bits downloaded from server $n$. 
Here, the expectation is taken over all realizations of $\mathbf{A}^{[\mathrm{W}]}_n$ (i.e., all realizations of $\mathbf{Q}^{[\mathrm{W}]}_n$ and all realizations of $\mathbf{X}_1,\dots,\mathbf{X}_K$). 

Our goal is to characterize the \emph{linear capacity} of MPIR (in terms of the parameters $N,K,D$) which is defined as the supremum of rates over all linear MPIR protocols. 


\section{Main Results}
In this section, we present our main results on the linear capacity of MPIR for all $K>1$ and $1<D\leq K$, and $N=D+1$. 

To simplify the notation, 
we define 
\begin{equation}\label{eq:ljmj}
l_j := \frac{\mathrm{lcm}(\binom{D}{j},D)}{D} \quad \text{and} \quad m_j:= \frac{Dl_j}{\binom{D}{j}}
\end{equation}
for all $1\leq j\leq D$, 
\begin{equation}\label{eq:L}
\mathrm{L}:=[l_1,\dots,l_D]^{\mathsf{T}},    
\end{equation}
and 
\vspace{-0.1cm}
\begin{equation}\label{eq:Lambda}
\mathrm{M} := 
\begin{bmatrix} 
l_1 & l_2 & \cdots & l_{D-2} & l_{D-1} & l_D\\
\frac{m_1}{m_2} & 0 & \cdots & 0 & 0 & 0\\
0 & \frac{m_2}{m_3} & \cdots & 0 & 0 & 0\\
\vdots & \vdots & \ddots & \vdots & \vdots & \vdots\\
0 & 0 & \cdots & \frac{m_{D-2}}{m_{D-1}} & 0 & 0\\
0 & 0 & \cdots & 0 & \frac{m_{D-1}}{m_{D}} & 0\\
\end{bmatrix}. 
\end{equation} 
Note that $\mathrm{L}$ is a column-vector of length $D$, and $\mathrm{M}$ is a $D\times D$ matrix. 
Also, we define 
\begin{equation}\label{eq:FG}
\mathrm{F}^{\mathsf{T}}:=\mathrm{L}^{\mathsf{T}}\mathrm{M}^{K-D} \quad \text{and} \quad \mathrm{G}^{\mathsf{T}}:=\mathrm{L}^{\mathsf{T}}(\mathrm{I}+\mathrm{M})^{K-D},    
\end{equation}
where $\mathrm{I}$ is the $D\times D$ identity matrix, and   
${\mathrm{F}=[f_1,\dots,f_D]^{\mathsf{T}}}$ and ${\mathrm{G}=[g_1,\dots,g_D]^{\mathsf{T}}}$ are column-vectors of length $D$.

\begin{theorem}\label{thm:1}
For MPIR with $K$ messages and demand's size $D$, when the number of servers is $N=D+1$, the linear capacity is lower bounded by \begin{equation}\label{eq:R}
R:=\frac{D}{N-\max_{j\in [D]} \frac{f_j}{g_j}}, 
\end{equation} 
and upper bounded by 
\begin{equation}\label{eq:C1}
\left(1+\frac{K-D}{DN}\right)^{-1}    
\end{equation}
or 
\begin{equation}\label{eq:C2}
\left(\frac{1-1/N^{\lfloor {K}/{D}\rfloor}}{1-1/N}+\left(\frac{K}{D}-\left\lfloor\frac{K}{D}\right\rfloor\right)\frac{1}{N^{\lfloor{K}/{D}\rfloor}}\right)^{-1},    
\end{equation}
when $D\geq \frac{K}{2}$ or $D\leq \frac{K}{2}$, respectively. 
\end{theorem}

The upper bounds in~\eqref{eq:C1} and~\eqref{eq:C2}---which appear without proof---follow directly from the converse results of~\cite{BU2018} on the general capacity of MPIR (i.e., the supremum of rates over all MPIR protocols). 
This is simply because the linear capacity cannot exceed the general capacity, and hence any upper bound on the general capacity serves also as an upper bound on the linear capacity.  
To prove the lower bound, we propose a new MPIR scheme that takes a carefully-designed randomized approach for constructing the query to each server. 
Our scheme does not rely on the idea of subpacketization, and can be considered as a nontrivial generalization of the PIR scheme of~\cite{TSC2019} for demand's size $D=1$ and $N=2$ servers, to the MPIR setting with demand's size $D>1$ and $N=D+1$ servers.  

\begin{theorem}\label{thm:2}
For MPIR with $K$ messages and demand's size $D$ such that $D\mid K$, when the number of servers is ${N=D+1}$, the linear capacity is given by 
\vspace{-0.1cm}
\begin{equation}\label{eq:C}
C:=\frac{1-1/N}{1-1/N^{{K}/{D}}}.    
\end{equation}
\end{theorem}
\vspace{-0.1cm}
It is easy to show that when $D\mid K$, the capacity upper bounds~\eqref{eq:C1} and~\eqref{eq:C2} are both equal to $C$---defined in~\eqref{eq:C}.
To prove Theorem~\ref{thm:2}, we  show that the achievable rate $R$---defined in~\eqref{eq:R}---is also equal to $C$ when $D\mid K$.

\begin{remark}\label{rem:1}\normalfont
As shown in~\cite{BU2018}, when $D\mid K$, for arbitrary number of servers $N>1$, the capacity $C$ is achievable using an MPIR scheme that requires each message to be divided into $N^2$ or more sub-messages of equal size. 
The result of Theorem~\ref{thm:2} shows that when $D\mid K$, for $N=D+1$ servers, the capacity $C$ can also be achieved using an MPIR scheme that does not require any subpacketization.
This proves that the MPIR scheme of~\cite{BU2018} is not always optimal in terms of the required level of subpacketization. 
\end{remark}

\begin{remark}\label{rem:2}\normalfont
When ${D\nmid K}$ and ${D<\frac{K}{2}}$, it can be shown that the rate of our scheme is no less than that of the scheme of~\cite{BU2018}, and surprisingly, in some cases, our scheme can achieve a higher rate. 
This highlights some advantages of PIR schemes that maximize the expected download rate over PIR schemes that maximize the download rate per instance.   
When ${D\nmid K}$ and ${D>\frac{K}{2}}$, our scheme achieves a lower rate than the scheme of~\cite{BU2018}. 
\end{remark}

\section{Proof of Theorem~\ref{thm:1}}
In this section, we propose a linear MPIR scheme which is applicable for any number of messages $K$ and any demand's size $D$, when the number of servers is $N=D+1$. 
Our scheme achieves the rate $R$ defined in~\eqref{eq:R}, hence proving the lower bound in Theorem~\ref{thm:1} on the linear capacity of MPIR.  
\vspace{-0.125cm}
\subsection{Proposed Scheme}
The proposed scheme consists of three steps described as follows. 

\vspace{0.1cm}
{\bf Step 1:} Given the demand's index set $\mathrm{W}$, the user generates $N$ sets $\{\mathrm{S}^{[\mathrm{W}]}_n\}_{n\in [N]}$, where for each ${n\in [N]}$, $\mathrm{S}^{[\mathrm{W}]}_n$ is a subset of $[K]$. 
Given $\{\mathrm{S}^{[\mathrm{W}]}_n\}_{n\in [N]}$, the user generates $N$ vectors $\{\mathrm{C}^{[\mathrm{W}]}_n\}_{n\in [N]}$, where for each ${n\in [N]}$, $\mathrm{C}^{[\mathrm{W}]}_n$ is a vector of length $K$ with support ${\mathrm{S}^{[\mathrm{W}]}_n}$ and 
nonzero entries from $\mathbbmss{F}^{\times}_q$.
Then, the user randomly chooses a permutation ${\pi: [N]\rightarrow [N]}$, and sends the query $\mathrm{Q}^{[\mathrm{W}]}_{\pi(n)}:=\mathrm{C}^{[\mathrm{W}]}_{n}$ to server $\pi(n)$.
The process of generating the collections ${\{\mathrm{S}^{[\mathrm{W}]}_n\}}$ and ${\{\mathrm{C}^{[\mathrm{W}]}_n\}}$ is based on a randomized algorithm---depending on $\mathrm{W}$---as described below.

For each $\mathrm{W}$, the set of all feasible collections $\{\mathrm{S}^{[\mathrm{W}]}_n\}$---among which the user generates one collection in order to construct the queries $\{\mathrm{Q}^{[\mathrm{W}]}_n\}$---can be represented by a table, labeled by $\mathcal{S}^{[\mathrm{W}]}$, 
each of whose rows specifies one feasible collection $\{\mathrm{S}^{[\mathrm{W}]}_n\}$. 
In the following, we describe the structure of the table $\mathcal{S}^{[\mathrm{W}]}$. 

For any integer ${0\leq i\leq K-D}$, let 
\vspace{-0.1cm}
\begin{equation}\label{eq:ki}
k_i:=\binom{K-D}{i}    
\end{equation} 
and let ${\mathrm{R}^{(i)}_{1},\dots,\mathrm{R}^{(i)}_{k_i}}$ be the collection (in an arbitrary but fixed order) of all $i$-subsets of ${[K]\setminus \mathrm{W}}$. 

Fix an arbitrary ${w\in \mathrm{W}}$.
For any integer ${1\leq j\leq D}$, let $l_j$ be as defined in~\eqref{eq:ljmj}, and 
let ${\mathrm{T}^{(j)}_{1},\dots,\mathrm{T}^{(j)}_{l_j}}$ be a collection (in an arbitrary but fixed order) of $l_j$ arbitrarily chosen distinct $j$-subsets of $\mathrm{W}$ that contain the index $w$. 
Let ${\mathrm{W} = \{w_0,\dots,w_{D-1}\}}$. 

For any ${\mathrm{T}^{(j)}_{l} = \{w_{r_1},\dots,w_{r_j}\}\subseteq \mathrm{W}}$ and any integer ${1\leq h\leq D}$, let ${\mathrm{T}^{(j)}_{l,h} := \{w_{s_1},\dots,w_{s_j}\}}$,
where \[{s_t:= r_t +h -1 \pmod D}\] for all ${1\leq t\leq j}$.



The table $\mathcal{S}^{[\mathrm{W}]}$ is composed of several components described as follows:
\begin{itemize}
    \item The table consists of ${K-D+1}$ sub-tables---one sub-table for each ${0\leq i\leq K-D}$; 
    \item Each sub-table $i$ consists of $k_i$ blocks---one block for each ${1\leq k\leq k_i}$; 
    \item Each block $k$ consists of $D$ sub-blocks---one sub-block for each ${1\leq j\leq D}$; and 
    \item Each sub-block $j$ contains $l_j$ rows---one row for each ${1\leq l\leq l_j}$.
\end{itemize}

It is easy to verify that the table $\mathcal{S}^{[\mathrm{W}]}$ consists of $2^{K-D}\sum_{j=1}^{D}l_j$ rows in total.
Indexing the rows of table $\mathcal{S}^{[\mathrm{W}]}$ by the tuples ${\{(i,k,j,l)\}}$, the row indexed by the tuple ${(i,k,j,l)}$, i.e., row $l$ in sub-block $j$ of block $k$ in sub-table $i$ of table $\mathcal{S}^{[\mathrm{W}]}$, is defined as follows:
\vspace{-0.125cm}
{\renewcommand{\arraystretch}{2.25}
\begin{table}[h!]
    \small
    \centering
    \begin{tabular}{|c|c|c|c|}
    \hline
     $\mathrm{S}^{[\mathrm{W}]}_1:=\mathrm{R}^{(i)}_{k}$ & $\mathrm{S}^{[\mathrm{W}]}_2:=\mathrm{R}^{(i)}_{k}\cup \mathrm{T}^{(j)}_{l,1}$ & \hspace{-0.2cm} $\cdots$ \hspace{-0.2cm} & $\mathrm{S}^{[\mathrm{W}]}_N:=\mathrm{R}^{(i)}_{k}\cup \mathrm{T}^{(j)}_{l,D}$ \\
     \hline
    \end{tabular}
\end{table}
}

\vspace{-0.3cm}
Note that each row consists of $N$ ($=D+1$) sets ${\mathrm{S}^{[\mathrm{W}]}_1,\dots,\mathrm{S}^{[\mathrm{W}]}_N}$. 
Also, note that by construction, the table $\mathcal{S}^{[\mathrm{W}]}$ may contain rows that are identical up to permutation.



To generate a collection $\{\mathrm{S}^{[\mathrm{W}]}_n\}_{n\in [N]}$, the user randomly selects a tuple $(i,k,j,l)$, where the probability of selecting the tuple $(i,k,j,l)$ is given by $P^{[\mathrm{W}]}_{i,k,j,l}$ for some ${0\leq P^{[\mathrm{W}]}_{i,k,j,l}\leq 1}$, and constructs the row corresponding to the tuple $(i,k,j,l)$ in the table $\mathcal{S}^{[\mathrm{W}]}$. 
Note that 
\vspace{-0.15cm}
\begin{equation}\label{eq:SumProb}
\sum_{i=0}^{K-D}\sum_{k=1}^{k_i}\sum_{j=1}^{D}\sum_{l=1}^{l_j} p^{[\mathrm{W}]}_{i,k,j,l} = 1. 
\end{equation}

The optimal choice of the probabilities $\{P^{[\mathrm{W}]}_{i,k,j,l}\}$ will be specified later in Section~\ref{sec:OptimalProb}.


\vspace{0cm}
Given the collection $\{\mathrm{S}^{[\mathrm{W}]}_n\}_{n\in [N]}$---which is determined by the tuple $(i,k,j,l)$ selected by the user, the user generates a collection of vectors $\{\mathrm{C}^{[\mathrm{W}]}_n\}_{n\in [N]}$, namely,  ${\mathrm{C}_1^{[\mathrm{W}]} := \mathrm{U}^{(i)}_{k}}$, $\mathrm{C}_2^{[\mathrm{W}]} :=  \mathrm{U}^{(i)}_{k} + \mathrm{V}^{(j)}_{l,1}$, $\dots$, $\mathrm{C}_N^{[\mathrm{W}]} :=  \mathrm{U}^{(i)}_{k} + \mathrm{V}^{(j)}_{l,D}$, 
where 
\begin{itemize}
    \item ${\mathrm{U}^{(i)}_{k}}$ is a vector of length $K$ whose support (i.e., the index set of the nonzero entries) is $\mathrm{R}^{(i)}_{k}$, and 
    the nonzero entries of ${\mathrm{U}^{(i)}_{k}}$ (i.e., the entries indexed by $\mathrm{R}^{(i)}_{k}$) are randomly chosen from $\mathbbmss{F}^{\times}_q$, and
    \item ${\mathrm{V}^{(j)}_{l,1},\dots,\mathrm{V}^{(j)}_{l,D}}$ are $D$ vectors of length $K$ whose supports are ${\mathrm{T}^{(j)}_{l,1},\dots,\mathrm{T}^{(j)}_{l,D}}$, respectively, and 
    the nonzero entries of ${\mathrm{V}^{(j)}_{l,1},\dots,\mathrm{V}^{(j)}_{l,D}}$ are randomly chosen from $\mathbbmss{F}^{\times}_q$ such that the $D\times K$ matrix formed by vertically stacking the vectors ${\mathrm{V}^{(j)}_{l,1},\dots,\mathrm{V}^{(j)}_{l,D}}$ is full rank. (Note that the existence of such vectors ${\mathrm{V}^{(j)}_{l,1},\dots,\mathrm{V}^{(j)}_{l,D}}$ is guaranteed for any field size $q>D$.)
\end{itemize}



\vspace{0.15cm}
{\bf Step 2:} Upon receiving the query  $\mathrm{Q}^{[\mathrm{W}]}_{\pi(n)} = \mathrm{C}^{[\mathrm{W}]}_{n}$, 
if $\mathrm{C}^{[\mathrm{W}]}_{n}$ is a nonzero vector, server $\pi(n)$ computes ${\mathrm{Y}_{n}= \sum_{i\in \mathrm{S}^{[\mathrm{W}]}_{n}} C^{[\mathrm{W}]}_{n,i} \mathrm{X}_{i}}$, and 
sends the answer ${\mathrm{A}^{[\mathrm{W}]}_{\pi(n)}:=\mathrm{Y}_{n}}$ back to the user, where 
$\mathrm{S}^{[\mathrm{W}]}_{n}$ is the support of the vector $\mathrm{C}^{[\mathrm{W}]}_{n}$, and $C^{[\mathrm{W}]}_{n,i}$ is the $i$th entry of the vector $\mathrm{C}^{[\mathrm{W}]}_{n}$.
Otherwise, if $\mathrm{C}^{[\mathrm{W}]}_{n}$ is a zero vector, server $\pi(n)$ does not send any answer back to the user, and the user sets $\mathrm{Y}_{n} = 0$.

\vspace{0.25cm}
{\bf Step 3:} Given the answers $\{\mathrm{A}^{[\mathrm{W}]}_{n}\}$, i.e., $\mathrm{Y}_1,\mathrm{Y}_2,\dots,\mathrm{Y}_N$, the user computes
${\mathrm{Z}_h := \mathrm{Y}_{h+1}-\mathrm{Y}_1}$ for all ${1\leq h\leq D}$. 
By construction, (i) $\mathrm{Z}_h$ is a linear combination of $j$ demand messages, namely, the demand messages indexed by $\mathrm{T}^{(j)}_{l,h}$, and 
(ii) the coefficient vectors pertaining to the linear combinations ${\mathrm{Z}_1,\dots,\mathrm{Z}_D}$, namely, the vectors ${\mathrm{V}^{(j)}_{l,1},\dots,\mathrm{V}^{(j)}_{l,D}}$, are linearly independent. 
Thus, the user can recover all $D$ demand messages by solving a system of linear equations.

\begin{example*}\normalfont
Consider a scenario in which $N=3$ servers store ${K=4}$ messages ${\mathrm{X}_1,\mathrm{X}_2,\mathrm{X}_3,\mathrm{X}_{4}}$, and a user wants to retrieve ${D=2}$ messages $X_1$ and $X_2$. 
That is, ${\mathrm{W} = \{1,2\}}$ and ${[K]\setminus \mathrm{W}} = \{3,4\}$.
Suppose that ${\mathrm{X}_1,\mathrm{X}_2,\mathrm{X}_3,\mathrm{X}_{4}\in \mathbbmss{F}_q}$ for some field size ${q>D=2}$, say, ${q=3}$. 
Note that for this example, ${0\leq i\leq K-D=2}$ and ${1\leq j\leq D=2}$, and hence, ${(k_0,k_1,k_2)=(1,2,1)}$ and ${(l_1,l_2)=(1,1)}$. 

Following the definitions, the index sets $\mathrm{R}^{(i)}_{k}$ for all ${0\leq i\leq 2}$ and ${1\leq k\leq k_i}$, and the index sets $\mathrm{T}^{(j)}_{l,h}$ for all ${1\leq j\leq 2}$, ${1\leq l\leq l_j}$ and ${1\leq h\leq D=2}$, are defined as
\begin{equation*}
{\renewcommand{\arraystretch}{1.5}
\begin{matrix}
\mathrm{R}^{(0)}_1 = \emptyset & \mathrm{R}^{(1)}_1 = \{3\} & \mathrm{R}^{(1)}_2 =\{4\} & \mathrm{R}^{(2)}_1 = \{3,4\}\\
\mathrm{T}^{(1)}_{1,1} = \{1\} & \mathrm{T}^{(1)}_{1,2} = \{2\} &  \mathrm{T}^{(2)}_{1,1} = \{1,2\} & \mathrm{T}^{(2)}_{1,2} = \{1,2\}
\end{matrix}
}
\end{equation*}

Using the index sets $\{\mathrm{R}^{(i)}_k\}$ and $\{\mathrm{T}^{(j)}_{l,h}\}$ defined as above, the table $\mathcal{S}^{[1,2]}$ can be constructed as follows (the first and last columns are not part of the table $\mathcal{S}^{[1,2]}$, and are presented only to show the index tuple corresponding to each row and the probability of selecting each row):


{\renewcommand{\arraystretch}{1.5}
\begin{table}[h]
    \centering
    \scalebox{1}{
    \begin{tabular}{|c|c|c|c|c|}
    \hline
    $(i,k,j,l)$ &
    $\mathrm{S}_{1}^{[1,2]}$ & $\mathrm{S}_{2}^{[1,2]}$ & $\mathrm{S}_{3}^{[1,2]}$  &   $P^{[1,2]}_{i,k,j,l}$ \\  
    \hline
     $(0,1,1,1)$ &  $\emptyset$ &   $\emptyset \cup \{1\}$ & $\emptyset \cup \{2\}$  &$\frac{1}{4}$ \\
    \hline
     $(0,1,2,1)$  & $\emptyset$ &   $\emptyset \cup \{1,2\}$  &  $\emptyset \cup \{1,2\}$  & $\frac{1}{12}$\\
    \hline
     $(1,1,1,1)$ &  $\{3\}$ &   $\{3\} \cup \{1\}$ & $\{3\} \cup \{2\}$  & $\frac{1}{6}$  \\
    \hline
    $(1,1,2,1)$  &  $\{3\}$ &   $\{3\} \cup \{1,2\}$  &  $\{3\} \cup \{1,2\}$  & $\frac{1}{12}$\\
    \hline
     $(1,2,1,1)$  &  $\{4\}$ &   $\{4\} \cup \{1\}$ & $\{4\} \cup \{2\}$   & $\frac{1}{6}$ \\
    \hline
    $(1,2,2,1)$  &  $\{4\}$ &   $\{4\} \cup \{1,2\}$  &  $\{4\} \cup \{1,2\}$  & $\frac{1}{12}$\\
    \hline
    $(2,1,1,1)$  &   $\{3,4\}$ &   $\{3,4\} \cup \{1\}$ & $\{3,4\} \cup \{2\}$  & $\frac{1}{6}$  \\
    \hline
    $(2,1,2,1)$ &   $\{3,4\}$ &   $\{3,4\} \cup \{1,2\}$  &  $\{3,4\} \cup \{1,2\}$  & $0$\\
    \hline
    \end{tabular}}
    \label{example}
\end{table}
}
\vspace{-0.25cm}
Suppose that the user selects the row indexed by the tuple ${(i,k,j,l) = (2,1,1,1)}$ with probability $\frac{1}{6}$.
Note that this row corresponds to the index sets $\mathrm{S}^{[1,2]}_{1} = \{3,4\}$, $\mathrm{S}^{[1,2]}_{2} = \{1,3,4\}$, and $\mathrm{S}^{[1,2]}_{3} = \{2,3,4\}$. 
Given these index sets, the user generates the vectors $\mathrm{C}^{[1,2]}_1 = \mathrm{U}^{(2)}_1$, $\mathrm{C}^{[1,2]}_2 = \mathrm{U}^{(2)}_1+ \mathrm{V}^{(1)}_{1,1}$, and $\mathrm{C}^{[1,2]}_3 = \mathrm{U}^{(2)}_1+ \mathrm{V}^{(1)}_{1,2}$, where 
\begin{itemize}
    \item $\mathrm{U}^{(2)}_1$ is a randomly chosen vector of length $K=4$ with support $\mathrm{S}^{[1,2]}_{1} = \{3,4\}$ and nonzero entries from ${\mathbbmss{F}^{\times}_3 = 1,2}$, say, ${\mathrm{U}^{(2)}_1 = [0,0,1,2]}$, and
    \item $\mathrm{V}_{1,1}^{(1)}$ and $\mathrm{V}_{1,2}^{(1)}$ are two randomly chosen linearly-independent vectors of length $K=4$ with supports $\mathrm{T}_{1,1}^{(1)} = \{1\}$ and $\mathrm{T}_{1,2}^{(1)} = \{2\}$, respectively, and nonzero entries from ${\mathbbmss{F}^{\times}_3 = \{1,2\}}$, say, ${\mathrm{V}_{1,1}^{(1)} = [2,0,0,0]}$ and ${\mathrm{V}_{1,2}^{(1)} = [0,1,0,0]}$.  
\end{itemize}
Then, the user sends the vectors  $\mathrm{C}^{[1,2]}_1 = [0,0,1,2]$, $\mathrm{C}^{[1,2]}_2 = [2,0,1,2]$, and $\mathrm{C}^{[1,2]}_3 = [0,1,1,2]$ to servers $\pi(1),\pi(2),\pi(3)$, respectively, for a randomly chosen permutation ${\pi: [3]\rightarrow [3]}$. 

Each server $\pi(n)$ computes a linear combination $\mathrm{Y}_{n}$ of messages $\mathrm{X}_1,\mathrm{X}_2,\mathrm{X}_3,\mathrm{X}_4$ with combination coefficients specified by the entries of the vector $\mathrm{C}^{[1,2]}_{n}$, and sends $\mathrm{Y}_n$ back to the user. 
For this example, ${\mathrm{Y}_1 = \mathrm{X}_3+\mathrm{X}_4}$, ${\mathrm{Y}_2 = 2\mathrm{X}_1+\mathrm{X}_3+\mathrm{X}_4}$, and ${\mathrm{Y}_3 = \mathrm{X}_2+\mathrm{X}_3+\mathrm{X}_4}$. 

Given $\mathrm{Y_1},\mathrm{Y_2},\mathrm{Y_3}$, the user subtracts off the contribution of $\mathrm{Y}_1$ from $\mathrm{Y}_2$ and $\mathrm{Y}_3$, and computes $\mathrm{Z}_1 = \mathrm{Y}_2-\mathrm{Y}_1 = 2\mathrm{X}_1$ and $\mathrm{Z}_2 = \mathrm{Y}_3-\mathrm{Y}_1 = \mathrm{X}_2$. Then, the user recovers the demand messages $\mathrm{X}_1$ and $\mathrm{X}_2$ from $\mathrm{Z}_1$ and $\mathrm{Z}_2$. 


Let ${\mathrm{W}_1 = \{1,2\}}$, ${\mathrm{W}_2 = \{1,3\}}$, ${\mathrm{W}_3 = \{1,4\}}$, ${\mathrm{W}_4 = \{2,3\}}$, ${\mathrm{W}_5 = \{2,4\}}$, ${\mathrm{W}_6 = \{3,4\}}$. 
To show that the privacy condition is satisfied, we need to show that for each ${n\in [3]}$, ${\mathbb{P}(\mathbf{W} = \mathrm{W}_m|\mathbf{Q}^{[\mathbf{W}]}_{\boldsymbol{\pi}(n)} = \mathrm{C}^{[1,2]}_{n})}$ is the same for all ${m\in [6]}$. 
Note that the support of $\mathrm{C}^{[1,2]}_n$ is $\mathrm{S}^{[1,2]}_n$, and the nonzero entries of $\mathrm{C}^{[1,2]}_n$ are chosen at random.  
Thus, it suffices to show that for each ${n\in [3]}$, 
${\mathbb{P}(\mathbf{W} = \mathrm{W}_m|\mathbf{S}^{[\mathbf{W}]}_{\boldsymbol{\pi}(n)} = \mathrm{S}^{[1,2]}_{n})}$ is the same for all ${m\in [6]}$, where $\mathbf{S}^{[\mathbf{W}]}_{\boldsymbol{\pi}(n)}$ is the support of $\mathbf{Q}^{[\mathbf{W}]}_{\boldsymbol{\pi}(n)}$.   
In the following, we show that this condition holds for $n=1$. 
The proof for $n=2$ and $n=3$ follows from similar arguments. 

Recall that in this example, ${\mathrm{S}^{[1,2]}_1 = \{3,4\}}$.  
To compute $\mathbb{P}(\mathbf{W} = \mathrm{W}_m|\mathbf{S}^{[\mathbf{W}]}_{\boldsymbol{\pi}(1)} = \{3,4\})$ for any $m\in [6]$, we need to consider all rows in tables $\mathcal{S}^{[\mathrm{W}_1]},\dots,\mathcal{S}^{[\mathrm{W}_6]}$ that contain the index set $\{3,4\}$, as shown below: 

\vspace{-0.125cm}
{\renewcommand{\arraystretch}{1.5}
\begin{table}[H]
    \centering
    \scalebox{1}{
    \begin{tabular}{|c|c|c|c|c|}
    \hline
    $(i,k,j,l)$ &
    $\mathrm{S}_{1}^{[1,2]}$ & $\mathrm{S}_{2}^{[1,2]}$ & $\mathrm{S}_{3}^{[1,2]}$  &   $P^{[1,2]}_{i,k,j,l}$ \\  
    \hline
     $(2,1,1,1)$  &  $\{3,4\}$ &   $\{3,4\} \cup \{1\}$ & $\{3,4\} \cup \{2\}$   & $\frac{1}{6}$ \\
    \hline
    \end{tabular}}
    \label{table13}
\end{table}
}
\vspace{-0.55cm}
{\renewcommand{\arraystretch}{1.5}
\begin{table}[H]
    \centering
    \scalebox{1}{
    \begin{tabular}{|c|c|c|c|c|}
    \hline
    $(i,k,j,l)$ &
    $\mathrm{S}_{1}^{[1,3]}$ & $\mathrm{S}_{2}^{[1,3]}$ & $\mathrm{S}_{3}^{[1,3]}$  &   $P^{[1,3]}_{i,k,j,l}$ \\  
    \hline
     $(1,2,1,1)$  &  $\{4\}$ &   $\{4\} \cup \{1\}$ & $\{4\} \cup \{3\}$   & $\frac{1}{6}$ \\
    \hline
    \end{tabular}}
    \label{table13}
\end{table}
}
\vspace{-0.55cm}
{\renewcommand{\arraystretch}{1.5}
\begin{table}[H]
    \centering
    \scalebox{1}{
    \begin{tabular}{|c|c|c|c|c|}
    \hline
    $(i,k,j,l)$ &
    $\mathrm{S}_{1}^{[1,4]}$ & $\mathrm{S}_{2}^{[1,4]}$ & $\mathrm{S}_{3}^{[1,4]}$  &   $P^{[1,4]}_{i,k,j,l}$ \\  
    \hline
     $(1,2,1,1)$  &  $\{3\}$ &   $\{3\} \cup \{1\}$ & $\{3\} \cup \{4\}$   & $\frac{1}{6}$ \\
    \hline
    \end{tabular}}
    \label{table14}
\end{table}
}
\vspace{-0.55cm}
{\renewcommand{\arraystretch}{1.5}
\begin{table}[H]
    \centering
    \scalebox{1}{
    \begin{tabular}{|c|c|c|c|c|}
    \hline
    $(i,k,j,l)$ &
    $\mathrm{S}_{1}^{[2,3]}$ & $\mathrm{S}_{2}^{[2,3]}$ & $\mathrm{S}_{3}^{[2,3]}$  &   $P^{[2,3]}_{i,k,j,l}$ \\  
    \hline
     $(1,2,1,1)$  &  $\{4\}$ &   $\{4\} \cup \{2\}$ & $\{4\} \cup \{3\}$   & $\frac{1}{6}$ \\
    \hline
    \end{tabular}}
    \label{table23}
\end{table}
}
\vspace{-0.55cm}
{\renewcommand{\arraystretch}{1.5}
\begin{table}[H]
    \centering
    \scalebox{1}{
    \begin{tabular}{|c|c|c|c|c|}
    \hline
    $(i,k,j,l)$ &
    $\mathrm{S}_{1}^{[2,4]}$ & $\mathrm{S}_{2}^{[2,4]}$ & $\mathrm{S}_{3}^{[2,4]}$  &   $P^{[2,4]}_{i,k,j,l}$ \\  
    \hline
     $(1,2,1,1)$  &  $\{3\}$ &   $\{3\} \cup \{2\}$ & $\{3\} \cup \{4\}$   & $\frac{1}{6}$ \\
    \hline
    \end{tabular}}
    \label{table24}
\end{table}
}
\vspace{-0.55cm}
{\renewcommand{\arraystretch}{1.5}
\begin{table}[H]
    \centering
    \scalebox{1}{
    \begin{tabular}{|c|c|c|c|c|}
    \hline
    $(i,k,j,l)$ &
    $\mathrm{S}_{1}^{[3,4]}$ & $\mathrm{S}_{2}^{[3,4]}$ & $\mathrm{S}_{3}^{[3,4]}$  &   $P^{[3,4]}_{i,k,j,l}$ \\  
    \hline
     $(0,1,2,1)$  &  $\emptyset$ &   $\emptyset \cup \{3,4\}$ & $\emptyset \cup \{3,4\}$   & $\frac{1}{12}$ \\
    \hline
    \end{tabular}}
    \label{table34}
\end{table}
}
\vspace{-0.125cm}
By Bayes' rule, we can write 
\begin{align*}
&\mathbb{P}(\mathbf{W} = \mathrm{W}_m | \mathbf{S}^{[\mathbf{W}]}_{\boldsymbol{\pi}(1)} = \{3,4\}) \\
& = \frac{\mathbb{P}(\mathbf{S}^{[\mathbf{W}]}_{\boldsymbol{\pi}(1)} = \{3,4\}|\mathbf{W} = \mathrm{W}_m)\mathbb{P}(\mathbf{W} = \mathrm{W}_m)}{\mathbb{P}(\mathbf{S}^{[\mathbf{W}]}_{\boldsymbol{\pi}(1)} = \{3,4\})}. 
\end{align*} 
Note that ${\mathbb{P}(\mathbf{W} = \mathrm{W}_m) = \frac{1}{6}}$ is the same for all ${m\in [6]}$, and ${\mathbb{P}(\mathbf{S}^{[\mathbf{W}]}_{\boldsymbol{\pi}(1)} = \{3,4\})}$ does not depend on $m$. 
Thus, we only need to show that ${\mathbb{P}(\mathbf{S}^{[\mathbf{W}]}_{\boldsymbol{\pi}(1)} = \{3,4\}|\mathbf{W} = \mathrm{W}_m)}$ is the same for all ${m\in [6]}$. 
Note that ${\mathbb{P}(\mathbf{S}^{[\mathbf{W}]}_{\boldsymbol{\pi}(1)} = \{3,4\}|\mathbf{W} = \mathrm{W}_1)} = \frac{1}{6}\times \frac{1}{3} = \frac{1}{18}$, where 
$\frac{1}{6}$ is the probability of the row in table $\mathcal{S}^{[\mathrm{W}_1]}$ that contains the index set $\{3,4\}$, and $\frac{1}{3}$ is the probability of $\boldsymbol{\pi}(1) = 1$. 
Similarly, it can be shown that ${\mathbb{P}(\mathbf{S}^{[\mathbf{W}]}_{\boldsymbol{\pi}(1)} = \{3,4\}|\mathbf{W} = \mathrm{W}_m)} =\frac{1}{18}$ for all $m=2,\dots,5$. 
Note also that ${\mathbb{P}(\mathbf{S}^{[\mathbf{W}]}_{\boldsymbol{\pi}(1)} = \{3,4\}|\mathbf{W} = \mathrm{W}_6)} = \frac{1}{12}\times \frac{2}{3} = \frac{1}{18}$, where $\frac{1}{12}$ is the probability of the row in table $\mathcal{S}^{[\mathrm{W}_6]}$ that contains the index set $\{3,4\}$, and $\frac{2}{3}$ is the probability of $\boldsymbol{\pi}(1) \in \{2,3\}$. 
Thus, ${\mathbb{P}(\mathbf{S}^{[\mathbf{W}]}_{\boldsymbol{\pi}(1)} = \{3,4\}|\mathbf{W} = \mathrm{W}_m)} = \frac{1}{18}$ for all ${m\in [6]}$.

\end{example*}

\subsection{Optimal Choice of Probabilities $\{P^{[\mathrm{W}]}_{i,k,j,l}\}$}\label{sec:OptimalProb}

For any $0\leq i\leq K-D$ and $1\leq j\leq D$, we set 
\[P^{[\mathrm{W}]}_{i,k,j,l}=P_{i,j}\] for all $1\leq k\leq k_i$, all $1\leq l\leq l_j$, and all $\mathrm{W}\in \mathbbmss{W}$, where the probabilities $\{P_{i,j}\}$---specified shortly---are chosen to maximize the rate of the scheme while satisfying the privacy condition. 
Note that for any given $i$ and $j$, all rows in sub-block $j$ of any block in sub-table $i$ of any table are assigned the same probability $P_{i,j}$ for some $0\leq P_{i,j}\leq 1$. 
Note also that the condition~\eqref{eq:SumProb} reduces to the following condition:
\begin{equation}\label{eq:NewSumProb}
\sum_{i=0}^{K-D}k_i\sum_{j=1}^{D} l_j P_{i,j} = 1.    
\end{equation}

\subsubsection*{Satisfying the Privacy Condition}
Fix an arbitrary ${n\in [N]}$, and let $\mathrm{C}$ be the query that the user sends to server $n$, and let $\mathrm{S}$ be the support of the vector $\mathrm{C}$.
To satisfy the privacy condition, ${\mathbb{P}(\mathbf{W}=\mathrm{W}|\mathbf{C}=\mathrm{C})}$ must not depend on $\mathrm{W}$, 
i.e., ${\mathbb{P}(\mathbf{W}=\mathrm{W}|\mathbf{C}=\mathrm{C})}$ must be the same for all ${\mathrm{W}\in \mathbbmss{W}}$. 
Since $\mathbf{W}$ is distributed uniformly over $\mathbbmss{W}$, it is easy to see that ${\mathbb{P}(\mathbf{W}=\mathrm{W}|\mathbf{C}=\mathrm{C})}$ is the same for all ${\mathrm{W}\in \mathbbmss{W}}$ 
so long as ${\mathbb{P}(\mathbf{C}=\mathrm{C}|\mathbf{W}=\mathrm{W})}$ is the same for all ${\mathrm{W}\in \mathbbmss{W}}$.  
Thus, we can write
\begin{align*}
&\mathbb{P}(\mathbf{C}=\mathrm{C}|\mathbf{W}=\mathrm{W})
= \mathbb{P}(\mathbf{S}=\mathrm{S} |\mathbf{W}=\mathrm{W})\times \mathbb{P}(\mathbf{C}=\mathrm{C}|\mathbf{W}=\mathrm{W},\mathbf{S}=\mathrm{S}).
\end{align*}
Since $\mathrm{S}$ is the support of the vector $\mathrm{C}$, and the nonzero entries of $\mathrm{C}$ are chosen at random, it readily follows that ${\mathbb{P}(\mathbf{C} = \mathrm{C}|\mathbf{W}=\mathrm{W},\mathbf{S}=\mathrm{S})}$ does not depend on $\mathrm{W}$. 
This implies that ${\mathbb{P}(\mathbf{C}=\mathrm{C}|\mathbf{W}=\mathrm{W})}$ is the same for all ${\mathrm{W}}$ so long as ${\mathbb{P}(\mathbf{S}=\mathrm{S}|\mathbf{W}=\mathrm{W})}$ is the same for all ${\mathrm{W}}$. 
Thus, the privacy condition is met so long as the probabilities $\{P_{i,j}\}$ are chosen such that ${\mathbb{P}(\mathbf{S}=\mathrm{S}|\mathbf{W}=\mathrm{W})}$ is the same for all ${\mathrm{W}}$.

Fix an arbitrary $\mathrm{W}\in \mathbbmss{W}$. 
Let ${\mathrm{R}:=\mathrm{S}\setminus \mathrm{W}}$ and ${\mathrm{T}:=\mathrm{S}\cap \mathrm{W}}$.
Note that \[{0\leq |\mathrm{R}|\leq \min\{|\mathrm{S}|,K-D\}} \quad \text{and} \quad {0\leq |\mathrm{T}|\leq \min\{|\mathrm{S}|,D\}}.\]

First, suppose that $|\mathrm{T}|=0$ (i.e.,  $\mathrm{T}=\emptyset$). 
In this case, it is easy to verify that $\mathrm{R}=\mathrm{R}^{(i)}_{k}$ for a fixed $i$ ($=|\mathrm{R}|$) and a fixed $k$ (depending on the ordering of the $i$-subsets of ${[K]\setminus \mathrm{W}}$ in the  collection ${\mathrm{R}^{(i)}_1,\dots,\mathrm{R}^{(i)}_{k_i}}$).
Note that $\mathrm{R}^{(i)}_{k}$ appears in the first column of all the rows indexed by the tuples $(i,k,j,l)$ for all $1\leq j\leq D$ and $1\leq l\leq l_j$, 
i.e., all rows in all sub-blocks of block $k$ in sub-table $i$ of table $\mathcal{S}^{[\mathrm{W}]}$.   
Thus, 
\begin{align}\label{eq:PTempty}
&\mathbb{P}(\mathbf{S}=\mathrm{S}|\mathbf{W}=\mathrm{W}) = \sum_{j=1}^{D}\sum_{l=1}^{l_j} P^{[\mathrm{W}]}_{i,k,j,l} \nonumber \\
& = \sum_{j=1}^{D}\sum_{l=1}^{l_j} P_{i,j} = \sum_{j=1}^{D} l_j P_{i,j} = \sum_{j=1}^{D} l_j P_{|\mathrm{R}|,j}.
\end{align}

Next, suppose that $|\mathrm{T}|\neq 0$ (i.e., $\mathrm{T}\neq\emptyset$). 
Similarly as in the previous case, 
it is easy to verify that $\mathrm{R}=\mathrm{R}^{(i)}_k$ for a fixed $i$ ($=|\mathrm{R}|$) and a fixed $k$ (depending on the ordering of the $i$-subsets of ${[K]\setminus \mathrm{W}}$ in the  collection ${\mathrm{R}^{(i)}_1,\dots,\mathrm{R}^{(i)}_{k_i}}$). 

In addition, it can be shown that $\mathrm{T}=\mathrm{T}^{(j)}_{l,h}$ for a fixed $j$ ($=|\mathrm{T}|$), and $m_j$ pairs ${(l,h)\in [l_j]\times [D]}$, where $m_j$ is as defined in~\eqref{eq:ljmj}.  
This is because: 
(i) for any ${1\leq j\leq D}$, there exist $\binom{D}{j}$ distinct $j$-subsets of $\mathrm{W}$, and 
(ii) for any ${0\leq i\leq K-D}$ and ${1\leq k\leq k_i}$, the $j$-subsets of $\mathrm{W}$ are distributed evenly (i.e., with the same multiplicity) in the last $D$ columns of the $l_j$ rows in sub-block $j$ of block $k$ in sub-table $i$ of table $\mathcal{S}^{[\mathrm{W}]}$. 
For any ${1\leq j\leq D}$ and ${1\leq l\leq l_j}$, let $m_{j,l}:=\left|\{h\in [D]: \mathrm{T}^{(j)}_{l,h} = \mathrm{T}\}\right|$. 
Note that $\sum_{l=1}^{l_j} m_{j,l} = m_j$.
Thus,
\begin{align}\label{eq:PTnonempty}
& \mathbb{P}(\mathbf{S}=\mathrm{S}|\mathbf{W}=\mathrm{W}) = \sum_{l=1}^{l_j} m_{j,l} P^{[\mathrm{W}]}_{i,k,j,l} \nonumber \\[-0.25cm]
& = \sum_{l=1}^{l_j} m_{j,l}P_{i,j} = m_{j}P_{i,j} = m_{|\mathrm{T}|}P_{|\mathrm{R}|,|\mathrm{T}|}. 
\end{align}

Recall that the probabilities $\{P_{i,j}\}$ must be chosen such that ${\mathbb{P}(\mathbf{S}=\mathrm{S}|\mathbf{W}=\mathrm{W})}$ is the same for all ${\mathrm{W}\in \mathbbmss{W}}$.

First, suppose that $\mathrm{S}= \emptyset$. 
In this case, for any ${\mathrm{W}\in \mathbbmss{W}}$,  ${\mathrm{S}\setminus \mathrm{W}=\emptyset}$ (i.e., ${|\mathrm{S}\setminus \mathrm{W}|=0}$) and ${\mathrm{S}\cap \mathrm{W} = \emptyset}$ (i.e., ${|\mathrm{S}\cap \mathrm{W}|=0}$), and hence, by~\eqref{eq:PTempty}, ${\mathbb{P}(\mathbf{S}=\mathrm{S}|\mathbf{W}=\mathrm{W})} = \sum_{j=1}^{D}l_j P_{|\mathrm{S}\setminus \mathrm{W}|,j} = \sum_{j=1}^{D}l_j P_{0,j}$, which does not depend on $\mathrm{W}$. 
Thus, in the case of $\mathrm{S}=\emptyset$,  ${\mathbb{P}(\mathbf{S}=\mathrm{S}|\mathbf{W}=\mathrm{W})}$ is the same for all ${\mathrm{W}\in \mathbbmss{W}}$, regardless of the choice of $\{P_{i,j}\}$.  

Next, suppose that $\mathrm{S}\neq \emptyset$. 
In this case, ${1\leq |\mathrm{S}|\leq K}$. 
For any $\mathrm{W}\in \mathbbmss{W}$, it is easy to see that 
${0\leq |\mathrm{S}\setminus \mathrm{W}|\leq \min\{|\mathrm{S}|,K-D\}}$ and
${0\leq |\mathrm{S}\cap \mathrm{W}|\leq  \min\{|\mathrm{S}|,D\}}$. 

For any ${0\leq t\leq\min\{|\mathrm{S}|,D\}}$, let $\mathbbmss{W}_{t}$ be the collection of all ${\mathrm{W}\in \mathbbmss{W}}$ such that ${|\mathrm{S}\cap \mathrm{W}|=t}$. 

We consider the following two cases separately: 
(i)~${1\leq |\mathrm{S}|\leq K-D}$, and 
(ii) ${K-D+1\leq |\mathrm{S}|\leq K}$. 

First, consider the case (i). 
In this case, ${\mathbbmss{W}_t\neq \emptyset}$ for all ${0\leq t\leq \min\{|\mathrm{S}|,D\}}$.
For all ${\mathrm{W}\in \mathbbmss{W}_0}$, ${|\mathrm{S}\cap \mathrm{W}| = 0}$ and ${|\mathrm{S}\setminus \mathrm{W}|=|\mathrm{S}|}$, and hence, 
\vspace{-0.125cm}
\begin{align}\label{eq:=0}
& \mathbb{P}(\mathbf{S} =\mathrm{S}|\mathbf{W}=\mathrm{W}) \stackrel{\footnotesize\text{\eqref{eq:PTempty}}}{=} \sum_{j=1}^{D}l_j P_{|\mathrm{S}|,j}. 
\end{align} 
In addition, for any ${1\leq t\leq \min\{|\mathrm{S}|,D\}}$, for all ${\mathrm{W}\in \mathbbmss{W}_t}$, ${|\mathrm{S}\cap \mathrm{W}|=t}$ and ${|\mathrm{S}\setminus \mathrm{W}|=|\mathrm{S}|-t}$, and hence, 
\begin{align}\label{eq:neq0}
& \mathbb{P}(\mathbf{S}=\mathrm{S}|\mathbf{W}=\mathrm{W}) \stackrel{\footnotesize\text{\eqref{eq:PTnonempty}}}{=} m_{|\mathrm{S}\cap \mathrm{W}|}P_{|\mathrm{S}\setminus \mathrm{W}|,|\mathrm{S}\cap \mathrm{W}|} = m_{t} P_{|\mathrm{S}|-t,t}. 
\end{align} 
According to~\eqref{eq:=0} and~\eqref{eq:neq0}, for any arbitrary $\mathrm{S}$ such that ${1\leq |\mathrm{S}|\leq K-D}$, it follows that ${\mathbb{P}(\mathbf{S}=\mathrm{S}|\mathbf{W}=\mathrm{W})}$ is the same for all ${\mathrm{W}\in \mathbbmss{W}}$ so long as
\begin{equation}\label{eq:orgCond11}
\sum_{j=1}^{D}l_j P_{|\mathrm{S}|,j} = m_{1} P_{|\mathrm{S}|-1,1}
\end{equation} 
and
\begin{equation}\label{eq:orgCond12}
m_{t} P_{|\mathrm{S}|-t,t} = m_{t+1} P_{|\mathrm{S}|-t-1,t+1}  
\end{equation}
for all ${1\leq t\leq \min\{|\mathrm{S}|,D\}-1}$.

Next, consider the case (ii).
In this case, it is easy to see that $\mathbbmss{W}_t=\emptyset$ for all ${0\leq t\leq |\mathrm{S}|-K+D-1}$, and $\mathbbmss{W}_t\neq \emptyset$ for all ${|\mathrm{S}|-K+D\leq t\leq \min\{|\mathrm{S}|,D\}}$.

For any ${|\mathrm{S}|-K+D\leq t\leq \min\{|\mathrm{S}|,D\}}$, 
for all ${\mathrm{W}\in \mathbbmss{W}_t}$, ${|\mathrm{S}\cap \mathrm{W}|=t}$ and ${|\mathrm{S}\setminus \mathrm{W}|=|\mathrm{S}|-t}$, and hence,
\begin{equation}\label{eq:PCaseii}
\mathbb{P}(\mathbf{S}=\mathrm{S}|\mathbf{W}=\mathrm{W}) \stackrel{\footnotesize\text{\eqref{eq:PTnonempty}}}{=} m_t P_{|\mathrm{S}|-t,t}.
\end{equation}
According to~\eqref{eq:PCaseii}, for any arbitrary $\mathrm{S}$ such that ${K-D+1\leq |\mathrm{S}|\leq K}$, it follows that ${\mathbb{P}(\mathbf{S}=\mathrm{S}|\mathbf{W}=\mathrm{W})}$ is the same for all ${\mathrm{W}\in \mathbbmss{W}}$ so long as
\begin{equation}\label{eq:orgCond2}
m_{t} P_{|\mathrm{S}|-t,t} = m_{t+1} P_{|\mathrm{S}|-t-1,t+1}
\end{equation} for all ${|\mathrm{S}|-K+D\leq t\leq \min\{|\mathrm{S}|,D\}-1}$. 

By the above arguments, the privacy condition is satisfied so long as the conditions~\eqref{eq:orgCond11}, \eqref{eq:orgCond12} and~\eqref{eq:orgCond2} are met.  
By a series of simple change of variables, it can be shown that these three conditions are met simultaneously so long as the probabilities $\{P_{i,j}\}$ satisfy the following two conditions: 
\begin{equation}\label{eq:FinalCond1}
\sum_{j=1}^{D}l_j P_{i,j} = m_{1} P_{i-1,1}    
\end{equation} for all ${1\leq i\leq K-D}$, and
\begin{equation}\label{eq:FinalCond2}
m_{j} P_{i,j} = m_{j+1} P_{i-1,j+1}
\end{equation} for all ${1\leq i\leq K-D}$ and ${1\leq j\leq D-1}$. 
Rewriting in matrix form, it follows that both conditions~\eqref{eq:FinalCond1} and~\eqref{eq:FinalCond2} are met so long as 
\begin{equation}\label{eq:FinalCond}
\mathrm{P}_{i-1} = \mathrm{M}\mathrm{P}_i   
\end{equation} for all ${1\leq i\leq K-D}$, or equivalently, 
\begin{equation}\label{eq:FinalFinalCond}
\mathrm{P}_{i} = \mathrm{M}^{K-D-i}\mathrm{P}_{K-D} 
\end{equation} for all ${0\leq i\leq K-D}$, where  
\begin{equation}\label{eq:Pi}
\mathrm{P}_i := [P_{i,1},\dots,P_{i,D}]^{\mathsf{T}},   
\end{equation} 
and $\mathrm{M}$ is as defined in~\eqref{eq:Lambda}.

The condition~\eqref{eq:FinalFinalCond} implies that the probability vectors $\mathrm{P}_0,\dots,\mathrm{P}_{K-D-1}$ (i.e., the probabilities $P_{i,j}$ for all ${1\leq i\leq K-D-1}$ and ${1\leq j\leq D}$) 
are uniquely determined given the probability vector $\mathrm{P}_{K-D}$ (i.e., the probabilities ${P_{K-D,1},\dots,P_{K-D,D}}$).
However, the probability vector $\mathrm{P}_{K-D}$ cannot be chosen arbitrarily. 
This is because the probabilities $\{P_{i,j}\}$ must satisfy the condition~\eqref{eq:NewSumProb}, i.e.,
$\sum_{i=0}^{K-D}k_i\sum_{j=1}^{D} l_j P_{i,j} = 1$.
Note that ${\sum_{j=1}^{D} l_j P_{i,j} = \mathrm{L}^{\mathsf{T}}\mathrm{P}_i}$, where $\mathrm{L}$ and $\mathrm{P}_i$ are defined in~\eqref{eq:L} and~\eqref{eq:Pi}, respectively.
Thus by~\eqref{eq:FinalFinalCond}, we can rewrite the LHS of~\eqref{eq:NewSumProb} as
\begin{align*}
\sum_{i=0}^{K-D} k_i \mathrm{L}^{\mathsf{T}}\mathrm{P}_i & = \sum_{i=0}^{K-D} k_i \mathrm{L}^{\mathsf{T}}\mathrm{M}^{K-D-i}\mathrm{P}_{K-D} \\
& = \mathrm{L}^{\mathsf{T}}\left(\sum_{i=0}^{K-D} k_i \mathrm{M}^{K-D-i}\right)\mathrm{P}_{K-D}\\
& = \mathrm{L}^{\mathsf{T}} (\mathrm{I}+\mathrm{M})^{K-D} \mathrm{P}_{K-D},
\end{align*} noting that $k_i = \binom{K-D}{i}$. 
Thus, the condition~\eqref{eq:NewSumProb} can be rewritten in matrix form as
\begin{equation}\label{eq:FFFCond}
\mathrm{L}^{\mathsf{T}} (\mathrm{I}+\mathrm{M})^{K-D} \mathrm{P}_{K-D} = 1,    
\end{equation} which imposes a condition on the probability vector $\mathrm{P}_{K-D}$.

\vspace{0.25cm}
\subsubsection*{Maximizing the Rate of the Scheme}
To maximize the rate of the proposed scheme, we need to minimize the expected number of bits downloaded from all servers, i.e., ${\sum_{n=1}^{N} \mathbb{E}[L(\mathbf{A}^{[\mathrm{W}]}_n)] = \sum_{n=1}^{N} \mathbb{E}_{\mathbf{Q}}[\mathbb{E}_{\mathbf{X}}[L(\mathbf{A}^{[\mathrm{W}]}_n)]]}$, where the inner expectation $\mathbb{E}_{\mathbf{X}}$ is taken over all realizations of   ${\mathbf{X}_1,\dots,\mathbf{X}_{K}}$, and the outer expectation $\mathbb{E}_{\mathbf{Q}}$ is taken over all realizations of $\mathbf{Q}^{[\mathrm{W}]}_{n}$. 
Recall that  $L(\mathbf{A}^{[\mathrm{W}]}_n)$ is the number of bits downloaded from server $n$.

Fix an arbitrary ${n\in [N]}$. 
Note that either (i) server $n$ receives a zero vector as the query from the user and hence, server $n$ does not send any answer back to the user, or (ii) the answer of server $n$ to the user's query is a nontrivial linear combination of the messages.
In the case (i), ${\mathbb{E}_{\mathbf{X}}[L(\mathbf{A}^{[\mathrm{W}]}_n)]=0}$ since server $n$ does not send any answer back to the user, 
whereas in case (ii), ${\mathbb{E}_{\mathbf{X}}[L(\mathbf{A}^{[\mathrm{W}]}_n)]=B = m\log_2 q}$ since the messages ${\mathbf{X}_1,\dots,\mathbf{X}_{K}}$ are identically and uniformly distributed over $\mathbbmss{F}_q^m$, and any nontrivial linear combination of the messages ${\mathbf{X}_1,\dots,\mathbf{X}_{K}}$ is uniformly distributed over $\mathbbmss{F}_q^m$.

Since server $n$ does not send any answer back to the user if and only if ${\mathbf{Q}^{[\mathrm{W}]}_n = 0}$, then it follows that
\begin{align*}
\mathbb{E}[L(\mathbf{A}^{[\mathrm{W}]}_n)] & = \mathbb{E}_{\mathbf{Q}}[\mathbb{E}_{\mathbf{X}}[L(\mathbf{A}^{[\mathrm{W}]}_n)]] \\
& = \left(1-\mathbb{P}\left(\mathbf{Q}^{[\mathrm{W}]}_n = 0\right)\right)B.    
\end{align*}

Note that the probability that ${\mathbf{Q}^{[\mathrm{W}]}_n = 0}$ is given by $\frac{1}{N}\sum_{j=1}^{D}l_j P_{0,j}$, where $\sum_{j=1}^{D}l_j P_{0,j}$ is the probability that the user selects one of the rows whose first column is the empty set, and $\frac{1}{N}$ is the probability that $\pi(1)=n$ for a randomly chosen permutation $\pi$.
Thus, 
\begin{align*}
\mathbb{E}[L(\mathbf{A}^{[\mathrm{W}]}_n)] = \left(1- \frac{1}{N}\sum_{j=1}^{D}l_j P_{0,j}\right)B.
\end{align*}
By symmetry,  $\mathbb{E}[L(\mathbf{A}^{[\mathrm{W}]}_1)] = \dots = \mathbb{E}[L(\mathbf{A}^{[\mathrm{W}]}_N)]$.
Thus, 
\begin{equation}\label{eq:DC}
\sum_{n=1}^{N} \mathbb{E}[L(\mathbf{A}^{[\mathrm{W}]}_n)] = \left(N-\sum_{j=1}^{D}l_j P_{0,j}\right)B,    
\end{equation}
which implies that the rate of the proposed scheme is 
\begin{equation}\label{eq:Rate}
\frac{D}{N-\sum_{j=1}^{D}l_j P_{0,j}}. 
\end{equation}

Note that 
\begin{equation}\label{eq:ObjMatrix}
\sum_{j=1}^{D} l_j P_{0,j} = \mathrm{L}^{\mathsf{T}}\mathrm{P}_{0} \stackrel{\footnotesize\text{\eqref{eq:FinalFinalCond}}}{=} \mathrm{L}^{\mathsf{T}}\mathrm{M}^{K-D}\mathrm{P}_{K-D},
\end{equation} 
where $\mathrm{L}$, $\mathrm{M}$, and $\mathrm{P}_i$'s are defined in~\eqref{eq:L},~\eqref{eq:Lambda}, and~\eqref{eq:Pi}, respectively.
By substituting~\eqref{eq:ObjMatrix} in~\eqref{eq:DC}, 
\begin{equation}\label{eq:FinalDC}
\sum_{n=1}^{N} \mathbb{E}[L(\mathbf{A}^{[\mathrm{W}]}_n)] = \left(N-\mathrm{L}^{\mathsf{T}}\mathrm{M}^{K-D}\mathrm{P}_{K-D}\right)B.    
\end{equation}

According to~\eqref{eq:FinalDC}, the expected number of bits downloaded from all servers is minimized so long as $\mathrm{L}^{\mathsf{T}}\mathrm{M}^{K-D}\mathrm{P}_{K-D}$ is maximized. 
Recall that the probability vector $\mathrm{P}_{K-D}$ must also satisfy the condition~\eqref{eq:FFFCond}, i.e., 
$\mathrm{L}^{\mathsf{T}} (\mathrm{I}+\mathrm{M})^{K-D} \mathrm{P}_{K-D} = 1$.
Thus, the optimal choice of the probability vector $\mathrm{P}_{K-D}$ can be found by solving the following optimization problem: 
\begin{align}\label{eq:Opt1}
\mathrm{max} & \quad \mathrm{F}^{\mathsf{T}}\mathrm{P}_{K-D}\\
\mathrm{s.t.} & \quad \mathrm{G}^{\mathsf{T}} \mathrm{P}_{K-D} = 1\nonumber\\
& \quad \mathbf{0}\leq \mathrm{P}_{K-D}\leq \mathbf{1}\nonumber
\end{align} where $\mathrm{F}^{\mathsf{T}}=\mathrm{L}^{\mathsf{T}}\mathrm{M}^{K-D}$ and $\mathrm{G}^{\mathsf{T}}=\mathrm{L}^{\mathsf{T}}(\mathrm{I}+\mathrm{M})^{K-D}$ are defined as in~\eqref{eq:FG}, and  $\mathbf{0}$ and $\mathbf{1}$ are the all-zero and all-one column-vectors of length $D$, respectively. 
Note that given the optimal probability vector $\mathrm{P}_{K-D}$, the rest of the probability vectors $\mathrm{P}_0,\dots,\mathrm{P}_{K-D-1}$ can be simply found by using~\eqref{eq:FFFCond}. 

For each $j\in [D]$, let $f_j$ and $g_j$ be the $j$th entry of the vectors $\mathrm{F}$ and $\mathrm{G}$, respectively. 
Then, we can rewrite the optimization problem~\eqref{eq:Opt1} as follows:
\begin{align}\label{eq:Opt2}
\mathrm{max} & \quad \sum_{j=1}^{D} f_j P_{K-D,j}\\
\mathrm{s.t.} & \quad \sum_{j=1}^{D} g_j P_{K-D,j} =  1\nonumber\\
& \quad 0\leq P_{K-D,j}\leq 1 \quad  \forall 1\leq j\leq D\nonumber
\end{align} 
Solving the optimization problem~\eqref{eq:Opt2}, 
it is easy to verify that the optimal solution is given by ${P_{K-D,j^{*}} = {1}/{g_{j^{*}}}}$, 
where ${j^{*} = \argmax_{j\in [D]} {f_j}/{g_j}}$, and 
${P_{K-D,j} = 0}$ for all ${j\in [D]\setminus \{j^{*}\}}$. 
Accordingly, the optimal value of the objective function is given by ${{f_{j^{*}}}/{g_{j^{*}}}=\max_{j\in [D]}{f_j}/{g_j}}$. 

Replacing $\mathrm{L}^{\mathsf{T}}\mathrm{M}^{K-D}\mathrm{P}_{K-D}$ by ${f_{j^{*}}}/{g_{j^{*}}}$ in~\eqref{eq:FinalDC}, the expected number of bits downloaded from all servers is given by 
$\left(N-{f_{j^{*}}}/{g_{j^{*}}}\right)B$, 
and hence, the rate of the proposed scheme is given by
${D}/{(N-{f_{j^{*}}}/{g_{j^{*}}})}$, which is the same as $R$ defined in~\eqref{eq:R}. 
This completes the proof of the lower bound in Theorem~\ref{thm:1}. 

\section{Proof of Theorem~\ref{thm:2}}
In this section, we show that when $D\mid K$, say, $\frac{K}{D} = L$ for some integer $L\geq 1$, 
the rate $R$ defined in~\eqref{eq:R} is equal to
\begin{equation}\label{eq:SimCap1}
\frac{1-1/N}{1-1/N^{L}},    
\end{equation}
where $N=D+1$.
It is easy to verify that~\eqref{eq:SimCap1} can be written as 
\begin{equation}\label{eq:SimCap2}
\frac{D}{N-(D+1)^{1-L}}.
\end{equation}
Comparing~\eqref{eq:Rate} and~\eqref{eq:SimCap2}, we need to show that 
\begin{equation}\label{eq:ToShow}
\sum_{j=1}^{D} l_j P_{0,j} = (D+1)^{1-L}.    
\end{equation}

By combining the conditions~\eqref{eq:FinalCond1} and~\eqref{eq:FinalCond2}, it can be shown that 
\begin{equation}\label{eq:Rec1}
P_{i,j} = \sum_{h=1}^{D} \frac{l_h}{m_h} P_{i+h,j}    
\end{equation} for all ${0\leq i\leq K-2D}$ and ${1\leq j\leq D}$. 

Defining $O_{i,j}:=P_{K-D-i,j}$, we can rewrite~\eqref{eq:Rec1} as 
\begin{equation}\label{eq:Rec2}
O_{i,j} = \sum_{h=1}^{D} \frac{l_h}{m_h} O_{i-h,j}    
\end{equation} for all ${D\leq i\leq K-D}$ and ${1\leq j\leq D}$. 
It is easy to see that~\eqref{eq:Rec2} represents $D$ linear recurrence relations of order $D$ in variable $i$---one recurrence for each ${1\leq j\leq D}$. 
Note that the $D$ initial values for the $j$th recurrence relation are $O_{0,j},\dots,O_{D-1,j}$, and these $D$ recurrence relations may have different initial values.  
The characteristic equation for each of these recurrence relations is given by 
${\lambda^D - \sum_{h=1}^{D}\frac{l_h}{m_h}\lambda^{D-h} = 0}$,
which is equivalent to
${D\lambda^D - \sum_{h=1}^{D}\binom{D}{h}\lambda^{D-h} = 0}$,
noting that ${\frac{l_h}{m_h} = \frac{1}{D}\binom{D}{h}}$ for all ${1\leq h\leq D}$. 
By the binomial theorem, ${\sum_{h=1}^{D} \binom{D}{h}\lambda^{D-h} =  (1+\lambda)^D-\lambda^D}$, and 
hence, the characteristic equation can be further simplified as
${(D+1)\lambda^D - (1+\lambda)^{D} = 0}$,  
which is a polynomial equation of degree $D$. 
Let $\lambda_1,\dots,\lambda_D$ be the roots of the characteristic equation. 
Note that 
\begin{equation}\label{eq:CE3}
(D+1)\lambda^D_h = (1+\lambda_h)^{D}  
\end{equation} for all ${1\leq h\leq D}$. 
Then, the solution to the $j$th recurrence relation is given by 
\begin{equation}\label{eq:Qij}
O_{i,j} = \sum_{h=1}^{D}\alpha_{j,h}\lambda^{i}_h    
\end{equation} 
for all ${0\leq i\leq K-D}$, 
where ${\alpha_{j,1},\dots,\alpha_{j,D}}$ are constants---uniquely determined by the initial values ${O_{0,j},\dots,O_{D-1,j}}$. 

Since $O_{i,j} = P_{K-D-i,j}$ by definition, it readily follows from~\eqref{eq:Qij} that 
\begin{equation}\label{eq:Pij}
P_{i,j} = \sum_{h=1}^{D}\alpha_{j,h}\lambda^{K-D-i}_h    
\end{equation} 
for all ${0\leq i\leq K-D}$. 
By substituting~\eqref{eq:Pij} in~\eqref{eq:NewSumProb}, we can rewrite~\eqref{eq:NewSumProb} as
\begin{equation}\label{eq:New2SumProb}
\sum_{i=0}^{K-D}k_i\sum_{j=1}^{D} l_j \sum_{h=1}^{D} \alpha_{j,h}\lambda^{K-D-i}_{h} = 1. 
\end{equation} 
Note that the LHS of~\eqref{eq:New2SumProb} can be further simplified as
\begin{align}
& \sum_{i=0}^{K-D}k_i\sum_{j=1}^{D} l_j \sum_{h=1}^{D} \alpha_{j,h}\lambda^{K-D-i}_{h} \nonumber\\
& \quad = \sum_{j=1}^{D} l_j \sum_{h=1}^{D} \alpha_{j,h}\sum_{i=0}^{K-D}k_i\lambda^{K-D-i}_{h} \label{eq:Int1}\\   
& \quad  = \sum_{j=1}^{D} l_j \sum_{h=1}^{D} \alpha_{j,h}\sum_{i=0}^{K-D}k_i\lambda^{i}_{h} \label{eq:Int2}\\
& \quad = \sum_{j=1}^{D} l_j \sum_{h=1}^{D} \alpha_{j,h} (1+\lambda_h)^{K-D},  \label{eq:Int3}  
\end{align} where~\eqref{eq:Int1} follows from changing the order of the summations;
\eqref{eq:Int2} follows from a simple change of variable and the symmetry of the binomial coefficients $k_i$;
and~\eqref{eq:Int3} follows from the binomial theorem. 

By using~\eqref{eq:Int3}, we can rewrite~\eqref{eq:New2SumProb} as
\begin{equation}\label{eq:New3SumProb}
\sum_{j=1}^{D} l_j \sum_{h=1}^{D} \alpha_{j,h} (1+\lambda_h)^{K-D} = 1.
\end{equation}
By assumption, ${K-D = (L-1)D}$, and $L-1$ is a non-negative integer. 
Raising both sides of the equation~\eqref{eq:CE3} to the power of $L-1$, it is easy to see that
\begin{equation}\label{eq:CE4}
(D+1)^{L-1}\lambda^{K-D}_h = (1+\lambda_h)^{K-D}    
\end{equation} for all ${1\leq h\leq D}$. 
By combining~\eqref{eq:New3SumProb} and~\eqref{eq:CE4}, it then follows that
\begin{equation}\label{eq:New4SumProb}
\sum_{j=1}^{D} l_j \sum_{h=1}^{D} \alpha_{j,h} \lambda_h^{K-D} = (D+1)^{1-L}.
\end{equation}

Now, we can show~\eqref{eq:ToShow} as follows:
\begin{align}
\sum_{j=1}^{D} l_j P_{0,j} & = \sum_{j=1}^{D} l_j \sum_{h=1}^{D} \alpha_{j,h} \lambda^{K-D}_h \label{eq:Last1}\\
& = (D+1)^{1-L},\label{eq:Last2}
\end{align} 
where~\eqref{eq:Last1} follows from replacing $P_{0,j}$ using~\eqref{eq:Pij},
and~\eqref{eq:Last2} follows from~\eqref{eq:New4SumProb}. 
This completes the proof of Theorem~\ref{thm:2}.

\section{Numerical Results}
In this section, we numerically compare the rate of the proposed scheme and the rate of the MPIR scheme in~\cite{BU2018}, for different parameters $K,D$, namely, ${D\in \{2,3,4\}}$ and ${K\in \{D+1,\dots,D+7\}}$. 
Each table presents the achievable rate of both schemes, the upper bound on the capacity, and the gap between the capacity upper bound and the rate of the proposed scheme, for a fixed value of $D$ and different values of $K$.


\begin{table}[h]
    \label{tab:1}
    \caption{}\vspace{-0.125cm}
    \centering
    \scalebox{1}{
    \begin{tabular}{|c|c|c|c|c|c|c|c|}
    \hline
    $D=2$ & $K=3$  &$K=4$& $K=5$ & $K=6$ & $K=7$  &  $K=8$ &   $K=9$ \\  
    \hline
     Rate of the proposed scheme &  $\sfrac{5}{6} $ &   $\sfrac{3}{4} $ & $\sfrac{57}{80} $  & $\sfrac{9}{13} $  &  $\sfrac{639}{938} $  &  $\sfrac{27}{40} $ &  $\sfrac{795}{1184} $ \\
    \hline
     Rate of the scheme in \cite{BU2018} &  $\sfrac{6}{7} $ &   $\sfrac{3}{4} $ & $\sfrac{42}{59}$  & $ \sfrac{9}{13}$  &  $\sfrac{156}{229} $  &  $\sfrac{27}{40} $ &  $\sfrac{1216}{1811} $ \\
     \hline
     Upper bound on the capacity &  $\sfrac{6}{7} $ &   $\sfrac{3}{4} $ & $\sfrac{18}{25} $  & $\sfrac{9}{13} $  &  $\sfrac{54}{79} $  &  $\sfrac{27}{40} $ &  $\sfrac{162}{241} $ \\
     \hline
     \begin{tabular}{c}
     Gap between the capacity upper bound\\ and the rate of the proposed scheme
     \end{tabular}
      &  $\sfrac{1}{42} $ &   $0 $ & $\sfrac{3}{400} $  & $0 $  &  $\sfrac{29}{12567} $  &  $0 $ &  $\sfrac{14}{18755} $ \\
     \hline
    \end{tabular}}\vspace{-0.25cm}
\end{table}

\begin{table}[h]
    \label{tab:2}
    \caption{}\vspace{-0.125cm}
    \centering
    \scalebox{1}{
    \begin{tabular}{|c|c|c|c|c|c|c|c|}
    \hline
    $D=3$ & $K=4$  &$K=5$& $K=6$ & $K=7$ & $K=8$  &  $K=9$ &   $K=10$ \\  
    \hline
     Rate of the proposed scheme&  $\sfrac{9}{10} $ &   $\sfrac{5}{6} $ & $\sfrac{4}{5} $  & $\sfrac{552}{707} $  &  $\sfrac{876}{1139} $  &  $\sfrac{16}{21} $ &  $\sfrac{1727}{2280} $ \\
    \hline
     Rate of the scheme in \cite{BU2018} &  $\sfrac{12}{13} $ &   $\sfrac{6}{7} $ & $\sfrac{4}{5} $  & $ \sfrac{324}{415}$  &  $\sfrac{876}{1139} $  &  $\sfrac{16}{21} $ &  $\sfrac{1727}{2280} $ \\
     \hline
     Upper bound on the capacity &  $\sfrac{12}{13} $ &   $\sfrac{6}{7} $ & $\sfrac{4}{5} $  & $\sfrac{48}{61} $  &  $\sfrac{24}{31} $  &  $\sfrac{16}{21} $ &  $\sfrac{192}{253} $ \\
     \hline
     \begin{tabular}{c}
     Gap between the capacity upper bound\\ and the rate of the proposed scheme\end{tabular} &  $\sfrac{3}{130} $ &   $\sfrac{1}{42} $ & $0 $  & $\sfrac{25}{4084} $  &  $ \sfrac{31}{6081}$  &  $0 $ &  $ \sfrac{57}{39664}$ \\
     \hline
    \end{tabular}}\vspace{-0.25cm}
\end{table}

\begin{table}[h]
    \label{tab:3}
    \caption{}\vspace{-0.125cm}
    \centering
    \scalebox{1}{
    \begin{tabular}{|c|c|c|c|c|c|c|c|}
    \hline
    $D=4$ & $K=5$  &$K=6$& $K=7$ & $K=8$ & $K=9$  &  $K=10$ &   $K=11$ \\  
    \hline
     Rate of the proposed scheme &  $\sfrac{14}{15} $ &   $\sfrac{22}{25} $ & $\sfrac{132}{155} $  & $\sfrac{5}{6} $  &  $\sfrac{605}{736} $  &  $\sfrac{883}{1084} $ &  $\sfrac{1187}{1466} $ \\
    \hline
    Rate of the scheme in \cite{BU2018} &  $\sfrac{20}{21} $ &   $\sfrac{10}{11} $ & $\sfrac{20}{23} $  & $\sfrac{5}{6} $  &  $\sfrac{605}{736} $  &  $\sfrac{883}{1084} $ &  $\sfrac{953}{1177} $ \\
     \hline
     Upper bound on the capacity &  $\sfrac{20}{21} $ &   $\sfrac{10}{11} $ & $\sfrac{20}{23} $  & $ \sfrac{5}{6}$  &  $\sfrac{100}{121} $  &  $\sfrac{50}{61} $ &  $\sfrac{100}{123} $ \\
     \hline
     \begin{tabular}{c}
     Gap between the capacity upper bound\\ and the rate of the proposed scheme
     \end{tabular} &  $\sfrac{2}{105} $ &   $\sfrac{8}{275} $ & $\sfrac{64}{3565} $  & $0 $  &  $ \sfrac{24}{5411}$  &  $\sfrac{160}{31397} $ &  $\sfrac{28}{8429} $ \\
     \hline
    \end{tabular}}
\end{table}

As can be seen, for fixed $D$, when $D\mid K$, both schemes achieve the same rate---equal to the upper bound on the capacity (and hence, the gap is zero). 
When $D\nmid K$, for small values of $K$ such that $D>\frac{K}{2}$, the scheme of~\cite{BU2018} achieves the capacity upper-bound, whereas the proposed scheme achieves a lower rate (and hence, the gap is nonzero). 
When $D\nmid K$, for values of $K$ such that $D<\frac{K}{2}$, both schemes achieve a rate lower than the capacity upper bound (and hence the gap is nonzero), and the rate of the proposed scheme is no less than that of the scheme in~\cite{BU2018}.
Interestingly, in this case, for some values of $K,D$, e.g., $D=2$ and $K=5$, the proposed scheme achieves a higher rate than the scheme in~\cite{BU2018}. 
When $D\nmid K$ and $D<\frac{K}{2}$, it can also be seen that the gap between the capacity upper bound and the rate of the proposed scheme decreases as $K$ increases. 



\bibliographystyle{IEEEtran}
\bibliography{PIR_PC_Refs}

\end{document}